\documentclass[12pt]{article}
\usepackage{epsfig}
\usepackage{cite}
\usepackage{latexsym}
\usepackage{amstex,amssymb}

\newcommand{\pd}{\partial}
\newcommand{\M}{{\cal M}}                        
\newcommand{\g}{g_{ab}}                          
\newcommand{\h}{h_{ab}}                          
\newcommand{\R}{{\rm I\!\rm R}}                  
\newcommand{\N}{{\cal N}}                        
\newcommand{\bM}{{\bar \M}}                      
\newcommand{\bm}{\bar m}                         
\newcommand{\Z}{{\mathbb Z}}

\newcommand{\dd}{{\rm d}}
\newcommand{\F}{{\cal F}}
\newcommand{\D}{{D}}    
\newcommand{\B}{{B}}

\begin{document}

\begin{center}
{\Large{\bf Topology Change and Causal Continuity}} 
\vskip 2mm
{\large Fay Dowker ${}^a$ and Sumati Surya ${}^b$} 
\vskip 2mm

{\small{\it ${}^a$Blackett Laboratory, Imperial College of Science,
Technology and Medicine,\\ London SW7 2BZ, United Kingdom\\
email:dowker$@@$ic.ac.uk}}

{\small{\it  ${}^b$ IUCAA, Post Bag 4, Ganeshkhind,  Pune 411 007,
India\\email:ssurya$@@$iucaa.ernet.in}}

\end{center}

\begin{abstract}
The result that, for a scalar quantum field propagating on a ``trousers''
topology in $1+1$ dimensions, the crotch singularity is a source for an
infinite burst of energy has been used to argue against the occurrence of
topology change in quantum gravity. We draw attention to a conjecture due
to Sorkin that it may be the particular type of topology change involved in
the trousers transition that is problematic and that other topology changes
may not cause the same difficulties. The conjecture links the singular
behaviour to the existence of ``causal discontinuities'' in the spacetime
and relies on a classification of topology changes using Morse theory.  We
investigate various topology changing transitions, including the pair
production of black holes and of topological geons, in the light of these
ideas.
\end{abstract}

\newtheorem{thm}{Theorem}
\newtheorem{lemma}{Lemma}
\newtheorem{defn}{Definition}
\section{Introduction}

It is widely believed that any complete theory of quantum gravity must
incorporate topology change. Indeed, within the particle picture of quantum
gravity \cite{geons} the frozen topology framework for a generic spatial
3-manifold leads to the problem of spin-statistics violations and such wild
varieties of quantum sectors that it seems that a frozen topology is
unmaintainable \cite{uir}. There is one result, however, that has been
cited as counter-evidence for topology change: that of the singular
propagation of a quantum field on a trousers spacetime in $1+1$ dimensions
\cite{dewitt}\cite{dray}. We will see how it may be possible to incorporate
this result naturally in a framework which nevertheless allows topology
change in general.

The most natural way of accommodating topology changing processes in
quantum gravity is using the Sum-Over-Histories (SOH) approach, although
there have also been some efforts in this direction within the Hamiltonian
picture \cite{bal}.  We take a {\it history} in quantum gravity to be a
pair $(\M, g)$, where $\M$ is a smooth $n$-dimensional manifold and $g$ is
a time-oriented Lorentzian metric on $\M$. (Strictly, a history is a
geometry and only represented by $(\M, g)$.)  The amplitude for the
transition from an initial space (${V}_0, q_0$) to a final space $({V}_1,
q_1)$, where the $V_i$ are closed $(n-1)$-manifolds and the $q_i$ are
Riemannian $(n-1)$-metrics, receives contributions from all compact
interpolating histories $(\M,g)$, satisfying the boundary conditions $\pd
{\M} = {V}_i \amalg {V}_f \\ $, ${g}|_{V_{i,f}}=q_{i,f}$ where $\amalg$ denotes
disjoint union and $V_0$ and $V_1$ are initial and final spacelike
boundaries of $(\M,g)$.  We call the manifold $\M$ such that $\pd {\M} =
{V}_i \amalg {V}_f$ a {\it topological cobordism} and $(\M, g)$ a {\it
Lorentzian cobordism}.  We will say that a topological cobordism or
a history is {\it topology changing} if $\M$ is not a product 
$V_0 \times I$, where $I$ is the unit interval. 
We will use the terminology {\it topology changing
transition} to refer to the transition from $V_0$ to $V_1$
when  
$V_0$ and $V_1$ are not diffeomorphic, without reference to any
particular cobordism.

When $V_0$ and $V_1$ are not diffeomorphic, the existence of a topological
cobordism, $\M$, is equivalent to the equality of the Stiefel-Whitney
numbers of $V_0$ and $V_1$ and is not guaranteed in arbitrary
dimensions. If a topological cobordism does not exist we would certainly
conclude that that transition is forbidden.  In 3+1 and lower dimensions,
however, a topological cobordism always exists.  Then, given a topological
cobordism, $\M$, a Lorentzian cobordism based on $\M$ will exist iff
 \cite{reinhart}\cite{sorkin} (1) $n$ is even and $\chi(\M) =0$ or (2)
$n$ is odd and $\chi(V_0) = \chi(V_1)$.  In 3+1 dimensions, a topological
cobordism with $\chi(\M) =0$ always exists and thus any three-dimensional
$V_0$ and $V_1$ are Lorentz cobordant.

The theorem of Geroch  \cite{geroch}, extended to $n$-spacetime
dimensions, tells us that if a time oriented Lorentzian metric exists on a
topology changing topological cobordism $\M$ then that metric must contain
closed timelike curves.  We consider these to be a worse pathology
than the alternative which is to allow certain singularities in the metric
{\it i.e.}, to weaken the restriction that the metric be Lorentzian
everywhere, and which, following the proposal of Sorkin  \cite{forks},
is what we will choose to do in this paper.  The singularities which we
need to admit in order to be able to consider all possible topological
cobordisms are rather mild.  Given any topological cobordism $(\M; V_0,
V_1)$, there exists an almost everywhere Lorentzian metric $g$ on $\M$
which has singularities which take the form of degeneracies where the
metric vanishes at (finitely many) isolated points.  These degeneracies
each take one of $(n+1)$ standard forms described by Morse theory as we
shall relate.  Allowing such singular metrics seems natural in light of the
fact that within the path integral formulation, paths are not always
required to be smooth; in fact they are known to be distributional.
Moreover, such degeneracies are allowed within a vielbien formulation of
gravity. For a discussion of these points, see \cite{rafjorma}.
   
So, by allowing such mildly singular Lorentz cobordisms in the SOH no
topological cobordism is excluded and, in particular, every transition
in 3+1 dimensions is viable at this level of the kinematics. We will
refer to these cobordisms as ``Morse cobordisms.''  However it seems
that dynamically some Morse cobordisms may be more equal than others.
The $1+1$ dimensional case gives us an idea about a possible class of
``physically desirable'' histories.  For a massless scalar quantum
field on a fixed (flat) metric on the $1+1$ trousers topology there is
an infinite burst of energy from the crotch singularity that
propagates along the future light cone of the singularity
\cite{dewitt}, \cite{dray}. This tends to suggest that such a history
would be suppressed in a full SOH for $1+1$ quantum gravity.  By
contrast, the singular behaviour of a quantum field on the background
of a flat metric on the  $1+1$ ``yarmulke'' cobordism ({\it i.e.} a
hemisphere representing creation/destruction of an $S^1$ from/to
nothing)  is of a significantly different nature, in
the sense that when integrated over the future null directions the
stress-energy is finite \cite{guraf}. The singularity in the yarmulke
case is therefore effectively ``squelched'', while it propagates in
the trousers.  Indeed, in studying $1+1$ models of topology change,
the authors of \cite{rafjorma} have found that there is a suppression
of the trousers cobordism in the SOH and an enhancement by an equal
factor of the yarmulke cobordism (over the trivial cylinder) and
separate from the suppression due to the backgrounds not being
classical solutions.

What features of the trousers and yarmulke might account for the different
behaviour of quantum fields in these backgrounds?  A closer look shows that
in the Morse cobordism on the trousers manifold an observer encounters a
discontinuity in the volume of her causal past as she traverses from the
leg region into the body. Since such a discontinuity is absent in the
yarmulke topology and cylinder topologies, Sorkin has conjectured that
there may be an intimate connection between the discontinuity in the volume
of the causal past/future of an observer (a {\it causal discontinuity}) and
the physically undesirable infinite burst of energy for a scalar field
propagating in such a background. And then further, that this could signal
a suppression of the amplitude for a causally discontinuous spacetime in
the full SOH in quantum gravity.

The plan for this paper is the following. In the next section we
include a review of Morse theory and surgery theory, thus setting the
stage for our work. We find that whenever a component of the
universe is created from nothing, its initial spatial topology must be
that of a sphere. In section \ref{sec:discont} we state a conjecture
of Borde and Sorkin that relates causal discontinuities to the Morse
``type'' of a cobordism. In order to lend substance to this
conjecture, we work out the example of the trousers topology in $1+1$
dimensions which also generalizes to higher dimensions.  In section
\ref{sec:gentop} we present an argument, due to Sorkin \cite{rafpriv},
that {\it any} topology changing transition in $3+1$ dimensions can be
achieved by some causally continuous Morse cobordism, once the
Borde-Sorkin conjecture is assumed to hold.  We then examine certain
specific examples of topology changing topological cobordisms in the
following two sections.  The first is the $3+1$ black-hole pair
production instanton studied in \cite{garstrom} \cite{dgkt}.  We show
by direct construction that a causally continuous Morse metric exists
on the background manifold of the instanton which is further evidence
that that particular topology change is one with a finite amplitude.
This result generalises simply to higher dimensions even though the
exact instantons are not known.  The second class of cobordisms we
analyse is a set of manifolds which describe in a natural way the pair
production of topological geons in the particle picture of prime
manifolds \cite{geons}. We will show that, unfortunately, these
manifolds do not support causally continuous Morse metrics. We
summarise these results in the last section and discuss their
implications.
 
\section{ Morse theory and surgery}
\label{sec:morse}

Suppose $\M$ is an $n$-dimensional, compact, smooth, connected manifold
such that $\pd \M$ has two disjoint $(n-1)$-dimensional components, $V_0$
and $ V_1$, which are closed and correspond to the initial and final
boundaries of the spacetime, respectively. 

Any such $\M$ admits a {\it Morse function}  $f:\M \rightarrow [0,1]$,
with $f|_{V_i} = 0 $, $f|_{V_f} =1$ such that f possesses a set of critical
points $\{p_k\}$ ($ \pd_a f(p_k)=0$) which are nondegenerate (i.e. the Hessian
$\pd_a\pd_f$ at these points is invertible). It follows that the critical
points of $f$ are isolated and that because $\M$ is compact, there are
only a finite number of them.  

Using this Morse function and any Riemannian metric $\h$ on $\M$, we may
then construct an almost everywhere Lorentzian metric on $\M$ with a finite
number of isolated degeneracies,
\begin{equation} 
\label{eqn:metric}
        \g = \h (h^{cd} {\pd}_c f{\pd}_df) - \zeta {\pd}_af {\pd}_bf,
\end{equation} 
where the constant $\zeta >1$ \cite{rafjorma}. Clearly, $\g$ is degenerate
(zero) precisely at the critical points of $f$. We refer to these points as
``Morse singularities''. Expressing a metric on $\M$ in terms of its Morse
functions $f$ relates the latter to the causal structure of the spacetime
in an intimate manner, as we will see.

We now make the proposal that in the SOH, for the amplitude for a
topology changing process, for each topological cobordism, only
metrics that can be expressed in the form (\ref{eqn:metric}) ({\it
  i.e.} which can be constructed from some Morse function and some
Riemannian metric) will be included. We call such metrics ``Morse
metrics.''  Note that since a Riemannian metric and Morse function
always exist on a given topological cobordism, no cobordism is ruled
out of the SOH at this kinematical level.

A comment is in order here to relate this proposal to previous work on
Lorentzian topology change and Morse theory. In work by Yodzis
\cite{yodzis} the attitude was taken that the Morse singularities should
not be considered as part of spacetime, in other words that the Morse
points themselves were to be removed by sending them to infinity. In
contrast, here we are suggesting that the Morse points should remain as
part of the spacetime.  Amongst other things, this entails extending the
usual gravitational action to Morse metrics. This is discussed in detail
for 1+1 dimensions in \cite{rafjorma}.  Keeping the Morse points still
allows a well-defined causal structure even at the Morse points and hence a
well-defined causal ordering of all the spacetime points.  This is
something which ties in well with the idea that the fundamental underlying
structure is a causal set.

        Before proceeding any further, we briefly review some relevant
properties of Morse functions which we will employ later. We have utilised
references \cite{milnor},\cite{gauld},\cite{dubrovin},\cite{nash},
\cite{sen} extensively for this purpose.
 
\begin{lemma}[Morse Lemma]
\label{lemma:mlemma}
         If $p \in \M$ is a critical point of a Morse function $f: \M
        \rightarrow [0,1]$, then there exists local coordinates $x_1, x_2
        \cdots x_n$ in some neighbourhood of $p$ in terms of which $f$ is
        given, in that neighbourhood, by $f(x_1,...x_n)=c - x_1^2 - x_2^2
        \cdots - x_{\lambda}^2 + x_{\lambda +1}^2 \cdots + x_n^2$ for $0
        \leq \lambda \leq n$ and $c = const$. 
\end{lemma}

The number of negative signs $\lambda$ in the above expression is the
number of maxima of $f$ at the point $p$ and is referred to as the {\it
Morse index} of $f$ at $p$. For example, the height function on the $1+1$
yarmulke topology has index 0 at the bottom point, while that on its time
reversed counterpart has index 2. The height function on the trousers
topology on the other hand has a Morse point of index 1 at the crotch as
does its time reverse.



The {\it Morse number} of $\M$ on the other hand is defined to be the
minimum over all Morse functions $f :\M \rightarrow [0,1]$ of the number of
critical points of $f$. Thus, for example, although the cylinder topology
in $1+1$ dimensions {\it allows} Morse functions with any even number of
critical points, its Morse number is nevertheless zero. We then refer to a
topological cobordism of Morse number 0 as a {\it trivial} cobordism and
that with Morse number 1 as an {\it elementary} cobordism.

\begin{lemma}
        Any cobordism can be expressed as a composition of elementary
        cobordisms \cite{milnor}.
\end{lemma}

\begin{figure}[t]
\centerline{\epsfig{file=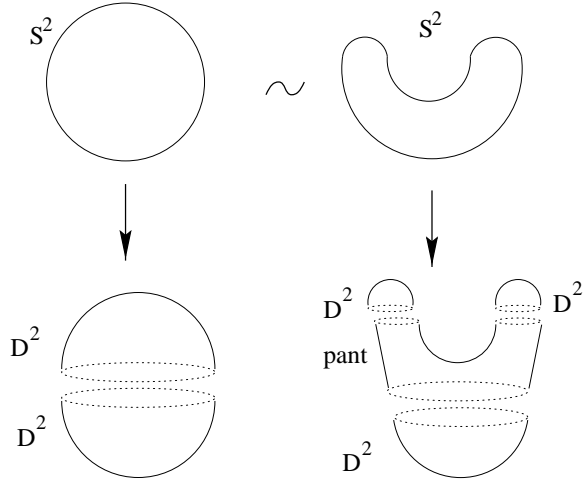,clip=,width=3in}}
\caption{Two ways of decomposing $S^2$ into elementary cobordisms.}
\label{fig:s2}
\end{figure}

This decomposition is however not unique, as can be seen in the
case of two dimensional closed universe $S^2$, shown in figure
(\ref{fig:s2}). Here we see that  $S^2$ could be decomposed into (a) two
elementary cobordisms,  yarmulke and its time reverse, or (b) into four
elementary cobordisms, namely, the yarmulke and an upside down trousers
topology with two time reversed yarmulkes, one capping each leg. 
Clearly, the causal structure of the two resulting histories is very different.


Before introducing surgery we define $D^k$ to be an open $k$ ball
and $B^k$ to be the closed $k$ ball (and $B^1 = I$).  

A {\it surgery} of type $\lambda$ on an $n-1$ dimensional manifold $V$ is
defined to be the following operation: Remove a thickened embedded
($\lambda -1$)-sphere, $S^{\lambda -1} \times
\D^{n-\lambda}$ from
$V$ and replace it with a thickened $({n-\lambda -1})$-sphere,
$S^{n-\lambda -1} \times \B^{\lambda}$ by identifying the boundaries using
a diffeomorphism, $ d: S^{\lambda -1} \times S^{n-\lambda -1} \rightarrow
S^{n-\lambda -1} \times S^{\lambda -1}$.

In performing a surgery, effectively, a $({\lambda -1})$-sphere is
``destroyed'' and an $({n-\lambda -1})$-sphere is ``created'' in this
process. 
We then have the following theorem which only
depends on surgery type.

\begin{thm}
\label{th:mil}
        If an $n-1$ dimensional manifold $V_1$ can be obtained from
        another $n-1$ dimensional manifold $V_0$ by a surgery of type
        $\lambda$, then $\exists$ an elementary cobordism $\M $,
        called the trace of a surgery, with
        boundary $V_0 \amalg V_1$ and a Morse function $f$ on $\M$, $f: M
        \rightarrow [0,1]$ which has exactly one Morse point of index
        $\lambda$ \cite{milnor}.
\end{thm}

\begin{figure}[t]
\centerline{\epsfig{file=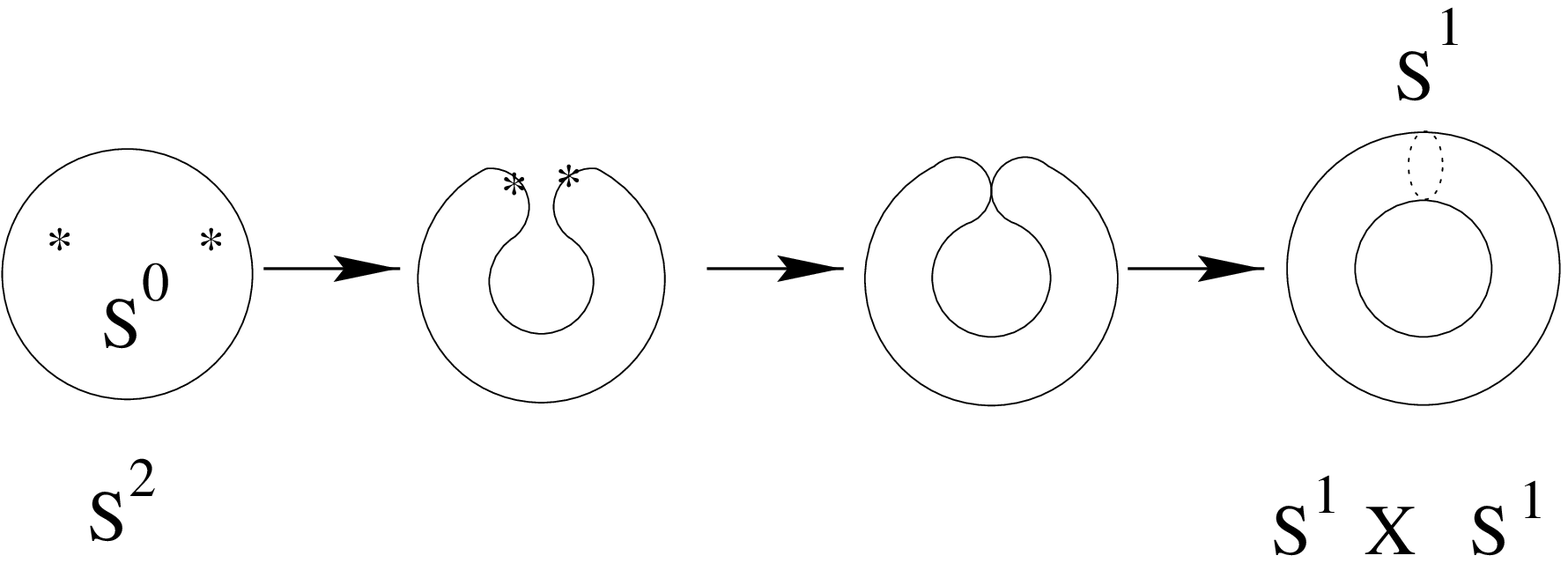,clip=,width=3in}}
\caption{``Tracing out'' a type $1$ surgery on $S^2$, whereby an $S^0$ is
destroyed and an $S^1$ is created to give the torus $S^1 \times S^1$.} 
\label{fig:worm}
\end{figure}

As an example, consider $V_0= S^2$ and $V_1= S^1 \times S^1$ or a
wormhole. Performing a type $1$ surgery on $S^2$ can result in the
manifold $S^1 \times S^1$, where an $S^0$ is ``destroyed'' and an
$S^1$ is ``created''.  Theorem \ref{th:mil} then says that $\exists$
an elementary cobordism $\M$ with boundary $S^2 \amalg S^1 \times S^1$
and a Morse function $f$ on $\M$ with a single critical point of index
$\lambda= 1$. The manifold $\M$ may be visualised as shown in figure
(\ref{fig:worm}).  We now explain how to construct the trace of a
general surgery.

A $\lambda$ surgery that turns $V_0$ into $V_1$  gives us  
an embedding $i: S^{\lambda -1} \rightarrow V_0$ and a neighbourhood,
$N$,
of that embedded sphere whose closure, $\bar N$, is
diffeomorphic to $S^{\lambda -1} \times B^{n-\lambda}$. Indeed, 
we have a diffeomorphism $d: \partial (\bar N) \rightarrow 
S^{\lambda-1} \times S^{n-\lambda-1}$, the ``surgery diffeomorphism.''
Now $S^{\lambda-1} \times S^{n-\lambda-1}$ is the boundary of
$S^{\lambda-1} \times B^{n-\lambda}$ and we can extend $d$ to a 
diffeomorphism, $\tilde d: \bar N \rightarrow  S^{\lambda-1} \times 
B^{n-\lambda}$ such that $\tilde d$ restricts to $d$ on the boundary.
$\tilde d$ is unique up to isotopy since $B^{n-\lambda}$ is topologically 
trivial.  

We construct the trace of the surgery by gluing together the two manifolds
$M_1 =V_0 \times I$ and $ M_2 = B^\lambda \times B^{n-\lambda}$ 
using a diffeomorphism from part of the 
boundary of one to part of the boundary of the other in the following way. 
$(\bar N, 1)$ is part of $\partial M_1$ and is diffeomorphic via $\tilde
d$ to  
$S^{\lambda -1} \times B^{n-\lambda}$ which is part of the boundary 
of $M_2$. We identify all points $x\in (\bar N, 1)$ and $\tilde d(x)$.
The resultant manifold clearly has one disjoint boundary component 
which is $V_0$. That the other boundary is diffeomorphic to $V_1$, 
{\it i.e.} the result of the surgery on $V_0$, takes a little more 
thought to see. Roughly speaking, in doing the gluing by 
$\tilde d$ we are eliminating $\bar N$ from $V_0$ and replacing it
with the rest of the boundary of $M_2$ (the complement of Im($\tilde d$)
in $\partial M_2$) {\it i.e.} $B^\lambda \times S^{n-\lambda -1}$
exactly as in the original surgery.  

\begin{figure}[t]
\centerline{\epsfig{file=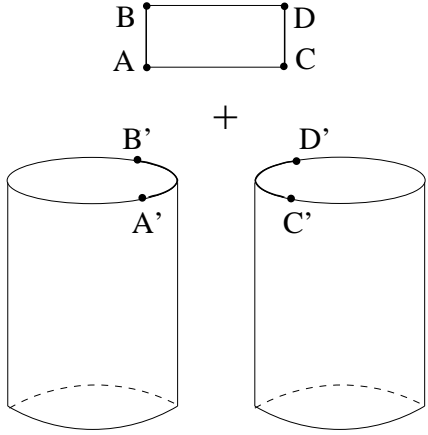,clip=,width=2.2in}}
\caption{Construction of the trace of a type $1$ surgery on $S^1\amalg
S^1$. The line segments $\vec{AB}$ and $\vec{CD}$ are identified with $\vec{A'B'}$ 
and $\vec{C'D'}$, respectively.}
\label{fig:trace}
\end{figure}

Figure (\ref{fig:trace}) is an example of the trace of
a type $1$ surgery on $S^1 \amalg S^1$, which is just the $1+1$ 
trousers topology. Here, $\bar N$ is the disjoint union of 
two line segments $\vec{AB}$ and $\vec{CD}$. 

Thus the trace of a surgery is a manifold with boundary with the property
that one  part of the boundary is the original manifold and the other part
of the boundary is the surgically altered manifold (up to diffeomorphisms).

\noindent{\bf Examples}

The $n$-dimensional yarmulke cobordism and its time reverse hold a
special place in our analysis since they are easy to characterise. If
$f: \M \rightarrow [0,1]$ has a single Morse point of index $0$ then
$\M$ is the trace of the surgery of type $0$ in which an $S^{-1}
\equiv \Phi$ is destroyed and an $S^{n-1}$ is created.  If $\M$ is
connected this implies that $\M \cong \B^n$.  In other words, a
cobordism can have a single index 0 point if and only if it is the
yarmulke. This means that when a component of the universe is created
from nothing (as opposed to being created by branching off from an
already existing universe) its initial topology must be that of a
sphere, no matter what the dimension: the big bang always results in
an initially spherical universe.  This might be thought of as a
``prediction'' of this way of treating topology change. A similar
argument for the time reversed case implies a connected cobordism can
have a single Morse point of index $n$ iff it is the time reversed
yarmulke and the universe must be topologically spherical before it
can finally disappear in a big crunch.
 
The trousers and its higher dimensional analogues are also important 
examples. There exists a Morse function on the $1+1$
trousers topology which possesses a single Morse point of index 1 and the
trousers is therefore the trace of a surgery of type $1$ in which an
embedded $S^0 \times \D^1$ is deleted from the initial $S^1\amalg S^1$ and 
replaced
with a $\B^1\times S^0$ to form a single  $S^1$.  
In $(n-1)+1$ dimensions, the higher dimensional trousers 
(the manifold $S^n$ with three open balls removed) for 
the process $S^{n-1} \amalg S^{n-1} \rightarrow S^{n-1}$ has an
index $1$
point and is the trace of a type $1$ surgery in which an
$S^0 \times \D^{n-2}$, {\it i.e.}, two balls, are removed and an
$S^{n-2} \times B^1$, or wormhole, added. In these processes, parts of
the universe which were spatially far apart suddenly become close
(in these cases the parts of the universe are originally
in disconnected components of the universe, but this isn't the 
defining characteristic of index 1 points). An index
$n-1$ 
point is the time reverse of this and corresponds to a type $n-1$ surgery in
which a wormhole is removed (or cut) and the ends ``capped off''
with two balls, so that neighbouring 
parts of the universe suddenly become distant.  

It seems intuitively clear from these examples 
that there is something causally peculiar 
about the index 1 and $n-1$ points and in the next section we give a 
precise statement of a conjecture that encapsulates this. 

\section{Causal Discontinuity}
\label{sec:discont}

\begin{figure}[t]
\centerline{\epsfig{file=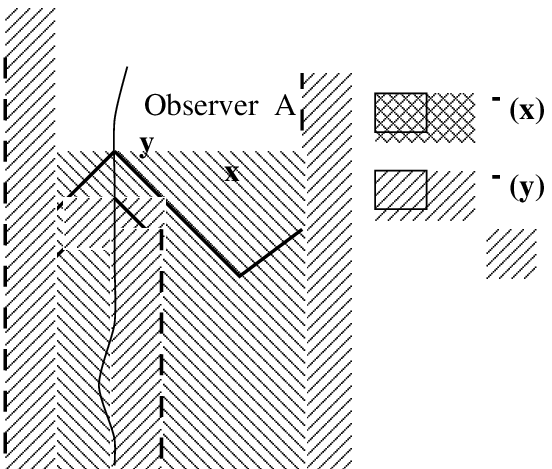,clip=,width=2.2in}}
\caption{Discontinuity in the causal past ${\rm I}^-$ of an observer A in
the trousers topology (the dashed lines are appropriately identified).}
\label{fig:cds}
\end{figure}

Borde and Sorkin have conjectured that $(\M, \g)$ contains a {\it causal
discontinuity} if and only if the Morse function $f$ contains an index 1 or
an index $n-1$ Morse point \cite{borde}. What do we mean by causal
discontinuity? There are many equivalent conditions 
for a Lorentzian spacetime to be causally discontinuous \cite{hawkingsachs} 
and we define a Morse metric to be causally discontinuous iff the 
spacetime minus the Morse points (which is Lorentzian) is. 
Roughly speaking, a causal discontinuity results in the causal past 
or future of a point in spacetime jumping discontinuously as the point
is continuously moved around. We see that behaviour in the 
1+1 trousers -- see figure (\ref{fig:cds}). Clearly the same kind of thing
will happen in the higher dimensional trousers, but not in the 
yarmulkes. Furthermore in the cases of index $\lambda \ne 1,n-1$, 
the spheres that are created and destroyed are all connected and so 
it seems that neighbouring parts of the universe remain close and 
distant ones remain far part.

\begin{figure}[t]
\centerline{\epsfig{file=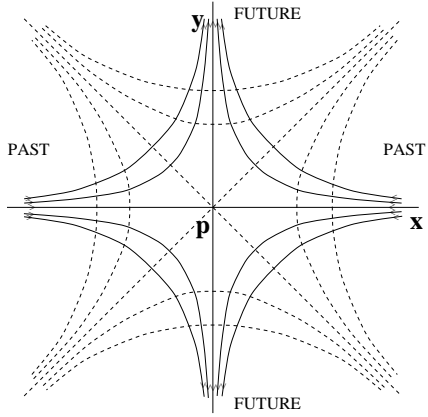,clip=,width=2.2in}}
\caption{The behaviour of the Morse function, $f$, around an index 1 point,
$p$, in $1+1$ dimensions. The solid lines are integral curves of $\xi^\mu =
h^{\mu\nu}\partial_\nu f$ with arrows in the direction of increasing $f$
and the dotted lines are surfaces of constant $f$.}
\label{fig:point}
\end{figure}

To  lend further plausibility to the conjecture  
we will work out an example, the index 1 point in 1+1
dimensions, in detail. Choose a neighbourhood of the Morse point 
$p$ in which the
Morse function has the standard form:

\begin{equation} 
\label{eqn:mfun}
{f(x,y) = f(p) - x^2 + y^2}
\end{equation} 
in terms of some local coordinates $(x,y)$. We take the flat
Riemannian metric
\begin{equation} 
\label{eqn:riem}
{ ds_R^2 = h_{\mu\nu} dx^\mu dx^\nu = dx^2 + dy^2}
\end{equation} 
Define the Morse metric $g_{\mu\nu}$ as in equation ~\ref{eqn:metric} with
$\zeta = 2$ and $\partial_\mu f = (-2x, 2y)$ to obtain

\begin{equation}        
\label{eqn:metex}
ds_L^2 = - 4(xdx-ydy)^2 + 4 (xdy + ydx)^2
\end{equation} 
This metric is actually flat since $ 2(xdx-ydy) = d(x^2 - y^2)$ and 
$2(xdy + ydx) = 2d(xy)$. In figure (\ref{fig:point}) we see that the hyperbolae
$xy = c$, $c$ constant, are the integral curves of the vector field 
$\xi^\mu = h^{\mu\nu} \partial_\nu f$  and the spatial ``surfaces''
of constant 
$f$ are the hyperbolae $x^2 - y^2 = d$, $d$ constant. 

\begin{figure}[t]
\centerline{\epsfig{file=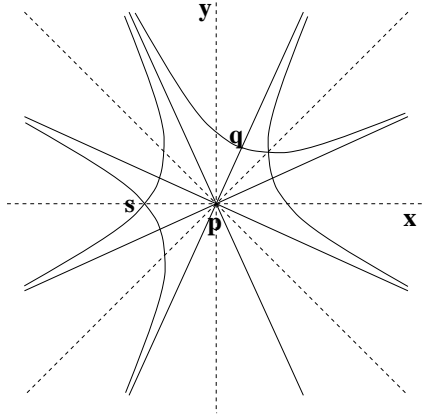,clip=,width=2.2in}}
\caption{Examples of null curves in a neighbourhood of $p$, all solid
lines. The straight lines are the past and future light cones of $p$. $q$
is a point on the future null cone of $p$.}
\label{fig:null}
\end{figure}

What are the null curves in the neighbourhood of $p$? We have $ds_L^2 = 0$
which implies
\begin{eqnarray} 
\label{eqn:nullc}
d(x^2 - y^2) & =& \pm 2 d(xy)\\
x^2 - y^2 &=& \pm 2xy + b
\end{eqnarray} 
The null curves that pass through $p$ are given by $b = 0$ so that there
are four solutions: ${y} =( \pm 1 \pm \sqrt{2})x$.  These are the
straight lines through $p$ at angles ${\pi\over 8}$, ${3\pi\over 8}$,
${5\pi\over 8}$, ${7\pi\over 8}$, to the $x$-axis.  These are the past and
future light ``cones'' of $p$. The null curves which don't pass through $p$
are given by the hyperbolae $x'y' = c'$ and ${x'}^2 - {y'}^2 = d'$ where
$(x',y')$ are rotated coordinates
\begin{eqnarray} 
x' &= &{\rm cos}{\pi \over 8} x + {\rm sin}{\pi\over 8}y\\ 
y'&  =& -{\rm sin}{\pi\over 8} x +  {\rm cos}{\pi \over 8}y. 
\end{eqnarray}  
Figure (\ref{fig:null}) shows a selection of null curves. In particular we
see the past and future light cones of point $s$ on the negative $x$-axis
and of a point $q$ on the future light cone of $p$. Using the results of
\cite{hawkingsachs} we can see that the spacetime around $p$ is not
causally continuous.  Indeed consider the point $q$
in figure (\ref{fig:null}). Then $\downarrow I^+(q) \ne I^-(q)$ where $I^+(q)$
($I^-(q)$) is the chronological future (past) of $q$ and $ \downarrow (S)$,
$S$ an open set, is the interior of the 
set of all points, $x$, for which there exists a
forward directed timelike curve from $x$ to every point in $S$. The point
$s$ is an element of $\downarrow I^+(q)$ but not $I^-(q)$.

The higher dimensional case can be similarly analysed. Now we have
\begin{equation}
\label{eqn:higher}
f(\vec{x}, \vec{y}) = f(p) - x_1^2 - \dots - x_\lambda^2 +
                                      y_1^2 + \dots + y_{n-\lambda}^2
\end{equation}
Take the Cartesian metric in the local coordinates and
let $r^2 = x_1^2 + \dots x_\lambda^2$ and $\rho^2 = y_1^2 + \dots y_{n-\lambda}^2$
so
\begin{equation}
\label{eqn:cart}
ds_R^2 = dr^2 + r^2 d\Omega_{\lambda-1}^2 + d\rho^2 + 
\rho^2 d\Omega_{n-\lambda-1}^2
\end{equation}
The Morse metric we construct from these and $\zeta = 2$ is 
\begin{eqnarray}
\label{eqn:highmor}
ds_L^2 & = & 4(r^2 + \rho^2)[r^2 d\Omega_{\lambda-1}^2 + \rho^2
         d\Omega_{n-\lambda-1}^2]\\  
       &  &+ 4(\rho dr + rd\rho)^2 - 4(rdr - \rho d\rho)^2
\end{eqnarray}  
This is not flat for $n\ge 3$. We can now solve $ds_L^2 = 0$ for a fixed
point on the $(\lambda-1)$-sphere and $(n-\lambda -1)$-sphere and find that
the past and future light cones of $p$ have base $S^{\lambda-1} \times
S^{n-\lambda -1}$. Note that this base is disconnected for $\lambda = 1$ or
$n-1$. The light cones of other points are more complicated to calculate
but a similar argument to that for the $1+1$ example shows that there is a
causal discontinuity for $\lambda = 1$ or $n-1$.

From now on we will assume that the Borde-Sorkin conjecture holds.
Thus, we can search for causally continuous histories on $\M$ by asking if
it admits {\it any} Morse function $f$ which has no index $1$ or $n-1$
critical points: a history corresponding to such an $f$ would be causally
continuous. If on the other hand, such an $f$ does {\it not} exist, i.e.,
all Morse functions on $\M$ have critical points of index either $1$ or
$n-1$, then $\M$ {\it does not} support causally continuous histories.  

We should remind ourselves 
that for a given Morse function $f$ on $\M$ the number
of index $\lambda$ critical points $m_\lambda$, is not a topological
invariant; in general different Morse functions will possess different sets
of critical points.  However there are lower bounds on the $m_\lambda$
depending on the homology type of $\M$. For the topological cobordism $(\M,
V_0, V_1)$ we have the Morse relation,
\begin{equation} 
\label{eqn:polynomial}
 {\sum}_\lambda (m_\lambda - {\beta}_\lambda(\M, V_0)) 
t^\lambda =(1+t)R(t),
\end{equation} 
where ${\beta}_\lambda(\M, V_0)$ are the Betti numbers of $\M$ relative to
$V_0$ and $R(t)$ is a polynomial in the variable $t$ which has positive
coefficients \cite{nash}\cite{sen}\cite{dubrovin}. Letting $t=-1$, we
immediately get the relative Euler characteristic of $\M$ in terms of the
Morse numbers,
\begin{equation} 
\label{eqn:morse}
 \chi(\M, V_0)= {\sum}_\lambda (-1)^{\lambda} m_\lambda.
\end{equation} 
Another consequence of (\ref{eqn:polynomial}) is, 
\begin{equation} 
\label{eqn:morsetwo}
 m_\lambda \geq {\beta}_\lambda(\M, V_0) \quad  \forall \lambda,
\end{equation} 
which places a lower bound on the $m_\lambda$.

\section{General topology change  in $n=4$}
\label{sec:gentop}

        As we have noted, in $n$-dimensions critical points of index $0$
and $n$ correspond to a big bang and big crunch, which allow causally
continuous histories. It is only for $n \geq 4$ that other types of
causally continuous histories can exist. For example, in $4$ dimensions,
elementary cobordisms with index $1$ or $3$ critical points correspond to
causally discontinuous histories while those of index $2$ are causally
continuous.

For $n=4$, we have already mentioned that any two $3$ manifolds $V_0$
and $V_1$ are cobordant, {\it i.e.}, $\exists$ a $4$ dimensional $\M$
such that $\pd \M = V_0 \amalg V_1$.  However, we can ask whether,
given a particular pair $\{V_0, V_1\}$, a cobordism $\M$ exists which
admits a causally continuous metric.  If not, then the Sorkin
conjecture would imply that the transition $V_0 \rightarrow V_1$ would
be suppressed. In other words, does a cobordism $\M$ exist which
admits a Morse function with no index $1$ or $3$ points? The answer to
this is supplied by a well known result in $3$ manifold theory, the
Lickorish-Wallace theorem, which states that any $3$-manifold $V_1$
can be obtained from any other $V_0$ by performing a series of type 2
surgeries on $V_0$ \cite{rafpriv}.  Thus, by Theorem {\bf 1} there
exists an interpolating cobordism $\M$ which is the trace of this
sequence of surgeries and which therefore admits a Morse function with
only index $2$ points, so that $\M$ admits a causally continuous
metric.

This result has the immediate consequence that even if the Sorkin and
Borde-Sorkin conjectures hold and causally discontinuous histories are
suppressed in the SOH, {\it no} topological transition $V_0 \rightarrow
V_1$ would be ruled out in 3+1 dimensions. Thus, in this sense, 
there is no ``causal'' obstruction to any transition $V_0 \rightarrow V_1$ 
in 3+1 dimensions, just as there is no topological (nor Lorentzian) obstruction
in  3+1 dimensions.

This is somewhat disappointing, however, since there are some transitions
that we might hope would be suppressed. An important example is the process
in which a single prime 3-manifold is produced. Quantised primes or
topological geons occur as particles in canonical quantum gravity similar
to the way skyrmions and other kinks appear in quantum field theory (see
\cite{geons} and section \ref{sec:geon} ).  We would therefore not expect
single geon production from the vacuum. However, the restriction of causal
continuity will not be enough to rule this out and we'll have to wait for
more dynamical arguments. This situation is in contrast to that for the
Kaluza-Klein monopole where there's a purely topological obstruction to the
existence of a cobordism for the creation of a single monopole
\cite{sorkin} (though that case is strictly not within the regime of our
discussion since the topology change involved is not local but changes the
boundary conditions at infinity).

This result, however, says nothing about the status of any
{\it particular} topological cobordism in the SOH. 
In other words, it may not be true that a given topological cobordism,
$\M$, admits a causally continuous Morse metric.

\section{Pair production of black holes}
\label{sec:blackholes}

The pair creation of black holes has been investigated by studying
Euclidean solutions of the equations of motion which satisfy the
appropriate boundary conditions for the solution to be an instanton for
false vacuum decay. One does not have to subscribe to the Euclidean
SOH approach to quantum gravity in order to believe that the
instanton calculations are sensible.  Indeed, we take the attitude that the
instantons are not ``physical'' but only machinery useful for approximately
calculating amplitudes \cite{forks} and that the functional integral is
actually over Morse metrics. The issue of whether quantum fields can
propagate in a non-singular way on these Morse geometries is therefore
relevant and the question arises as to whether causally continuous Morse
metrics can live on the instanton manifold.

The doubled instanton, or bounce, corresponding to the pair creation and
annihilation of non-extremal black holes has the topology $S^2 \times S^2
-{pt}$ \cite{garstrom}.  Let us compactify this to $S^2 \times S^2$.  The
fact that $S^2 \times S^2$ is closed implies that it will include at least
one universe creation and one universe destruction, corresponding to Morse
index $0$ and $4$ points, respectively.  This can be seen from the Betti
numbers, $\beta_0=\beta_4=1$, $\beta_1=\beta_3=0$ and $\beta_2 = 2$ so the
Morse inequalities imply that $m_0 \ge 1$ and $m_4\ge 1$.  Although
$\beta_1=\beta_3=0$ we cannot conclude that there exists a Morse function
that saturates the bounds of the inequalities (see the next section for
an example). We will prove that such a Morse function exists (with
$m_0=m_4=1$, $m_1=m_3 =0$ and $m_2=2$) by explicit construction on the
half-instanton, $S^2 \times \B^2$.

Let $(\theta, \phi)$ be standard polar coordinates on $S^2$ and $(r,\psi)$
polar coordinates on $\B^2$, where $\theta \in [0,\pi]$, $\phi \in [0,
2\pi]$, $0 \leq r \leq 1$ and $\psi \in [0, 2\pi]$. The boundary of $S^2
\times \B^2$ is $S^2\times S^1$ so that $S^2 \times \B^2$ corresponds to
the creation from nothing of an $S^2 \times S^1$ wormhole.

\begin{figure}[t]
\centerline{\epsfig{file=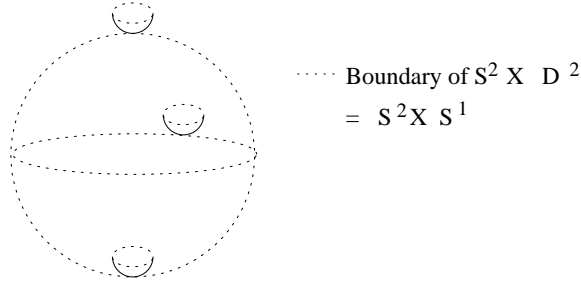,clip=,width=3in}}
\caption{The level surface $f^{-1}(1)$ is the boundary $S^2 \times S^1$
of $S^2 \times B^2$. }
\label{fig:boundary}
\end{figure}

Define  the function, 
\begin{equation} 
\label{eqn:eff}
f( \theta, \phi, r, \psi)= {1\over 3}(1+ r^2 + \cos
(1-r^2)\theta).
\end{equation} 
Now, $f:S^2 \times \B^2 \rightarrow [0,1]$. The level surface $f^{-1}(1)$
satisfies the condition $r=1$. This is easily seen to be the boundary $S^2
\times S^1$ of $S^2 \times \B^2$ (figure (\ref{fig:boundary})).  On the
other hand, the level surface $f^{-1}(0)$ satisfies the condition $r=0,
\theta=\pi$ which is a point on $S^2 \times B^2$.

We find the critical points of $f$ by noting that ${\pd}_rf = {2\over 3}r +
{2\over 3}r \theta \sin (1-r^2)\theta$ and ${\pd}_{\theta}f= -{1\over
3}(1-r^2) \sin (1-r^2) \theta$, while ${\pd}_{\phi}f={\pd}_{\psi}f=0$
everywhere. Thus, there are only two (and therefore isolated) critical
points of $f$, i.e., $p_1 = (r=0,\theta= \pi)$ and $p_2=(r= 0, \theta
=0)$ which are not on the boundary.  In
order to show the critical points are non-degenerate and 
to determine their indices we make use
of the Morse Lemma and rewrite $f$ in suitable local coordinate patches.

\vskip 3mm

\noindent {\bf  Near $p_1$}: At $p_1$, $f = 0$. In the neighbourhood of
$p_1$, we may write  $\theta = \pi - \epsilon$ where $\epsilon$ and $r$ are
both small and of the same order (note that the topology of this
neighbourhood is just $\B^2 \times \B^2$). Then, 
\begin{eqnarray}  
\label{eqn:nearpone}
\cos{(1-r^2)\theta}&\approx & \cos{(\pi - \epsilon)}\\
                   &\approx & -1 + {1\over 2} \epsilon^2.
\end{eqnarray}
and putting  $x_1 = {r\over \sqrt 3}\sin\psi$, $x_2 = {r\over \sqrt
3}\cos\psi$, $x_3 ={\epsilon \over \sqrt 6}\sin\phi$ and $x_4 ={\epsilon
\over \sqrt 6}\cos\phi$, we see that
\begin{equation} 
\label{eqn:fnearone} f\approx x_1^2 + x_2^2 + x_3^2 + x_4^2. 
\end{equation} 
Thus, $p_1$ is an index 0 point.

\vskip 3mm

\noindent {\bf Near $p_2$}: At $p_2$, $f = {2\over3}$. In the neighbourhood
of $p_2$, $r$ and $\theta$ are small and of the same order. Then,
\begin{equation} 
\label{eqn:effnear} 
f  \approx  {2\over 3} + {1\over 3} r^2 - {1\over 6} \theta^2, 
\end{equation} 
and using $y_1 ={\theta \over \sqrt 6}\sin\phi$ and $y_2 ={\theta \over
 \sqrt 6}\cos\phi$, $y_3 = {r\over \sqrt 3}\sin\psi$, $y_4 = {r\over \sqrt
 3}\cos\psi$, we see that
\begin{equation} 
f  \approx   {2\over 3} - y_1^2 - y_2^2 + y_3^2 + y_4^2.
\end{equation}  
So $p_2$ is an index 2 point.

\vskip 2mm

The existence of such a Morse function with two critical points, one of
index 0 and the other of index 2, shows that the black hole pair production
topology can support histories that are causally continuous. The index 0
point is the creation of an $S^3$ from nothing and the index 2 point is the
transition from $S^3$ to $S^2 \times S^1$. That this is means that a Morse
function with the same Morse points exists on the original non-compact
cobordism, half of $S^2 \times S^2 - \{{\rm point}\}$ was later shown in
\cite{dowgar}.  This result is evidence of consistency between the
conclusion that the existence of an instanton implies that the process has
a finite rate (approximated by $\tilde e^{-I}$ where $I$ is the Euclidean
action) and the idea that only causally continuous Morse histories
contribute to the SOH.

We note that a simple generalisation of the above Morse function shows that
the higher dimensional black hole pair creation-annihilation topological
cobordism $S^{n-2} \times \B^2$ admits a Morse function with one index 0
point and an index $(n-2)$ point and so supports histories that are
causally continuous for any dimension $n> 4$ (though the actual instanton
solution is unknown).  It is also interesting that there is another simple
cobordism for the transition from $S^3$ to $S^2 \times S^1$ which is $\B^3
\times S^1$ with an embedded open four-ball deleted.  This, however, by
virtue of the Morse inequalities, admits no Morse function without an index
1 point and so is causally discontinuous.  In some sense, this second
causally discontinuous process is the way one might naturally imagine a
wormhole forming: two distant regions of space coming ``close in
hyperspace'' and touching to form the wormhole. The index 2 cobordism for
creation of a wormhole is harder to visualise.

\section{Pair production of topological geons} 
\label{sec:geon}  

\begin{figure}[t]
\centerline{\epsfig{file=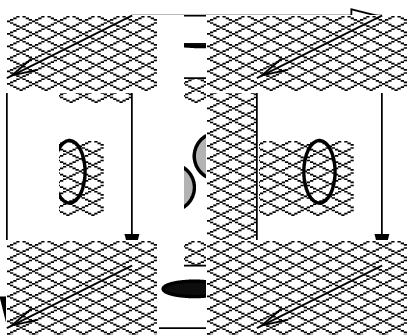,clip=,width=1.5in}}
\caption{A $T^3$ prime: the opposite sides of the cube are identified as
are the opposite edges.}
\label{fig:t3}
\end{figure}

Topological geons are particles that exist because of the non-trivial
topology of space. A geon is based on a prime 3-manifold, one which cannot
be divided further into non-trivial pieces by  embedded 2-spheres. One can
build a kinematical particle picture in quantum gravity whereby the geons
can be endowed with spin and statistics \cite{geons} \cite{amalfi}
\cite{fsor} \cite{uir}.  Every prime can be constructed from a solid
polyhedron by  identifying its boundary in some way -- it is
helpful in what follows to imagine the prime as a torus, $T^3$, so the
polyhedron is a solid cube and opposite faces are identified (figure
(\ref{fig:t3})).
To take the connected sum of a prime $P$ with any three-manifold $V$,
denoted $P\# V$, the (open) solid polyhedron is deleted from 
$V$ and the same identifications made on the resultant boundary. 
(The connected sum is also formed by
 removing open balls from each of two three-manifolds and identifying
the resulting $S^2$ boundaries.)

\begin{figure}[t]
\centerline{\epsfig{file=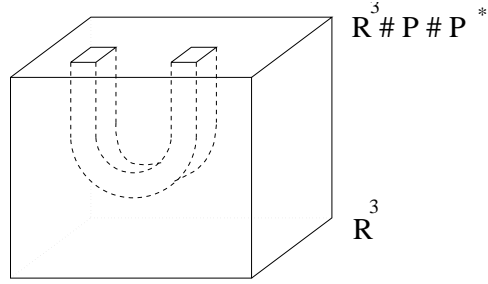,clip=,width=2.5in}}
\caption{A 2+1 dimensional representation of the U-tube cobordism for $\R^3 
\rightarrow \R^3 \#P \#P^*$. }
\label{fig:utube}
\end{figure}

A rather natural cobordism for pair-production of topological geons,
inspired by its Feynman-diagram likeness, is the ``U-tube'' \cite{higuchi}
\cite{fay}. Figure (\ref{fig:utube}) is a 2+1 sketch of this manifold which
is formed by removing a U-tube of solid polyhedral cross-section out of
$\R^3 \times I$ as shown and identifying the resulting boundaries in a
manner appropriate to the prime $P$. The initial boundary is $\R^3$ and the
final boundary is $\R^3 \#P\# P^*$ where $P^*$ denotes the chiral conjugate
(mirror image) of $P$. (In our example, $T^3$ is self-conjugate.)  Such a
U-tube cobordism was used to prove a spin-statistics correlation for
certain lens space topological geons (all of which are self-conjugate)
\cite{fay}. Moreover, the argument that certain proposed rules for
assigning quantum phases to different cobordisms would give a completely
general spin-statistics correlation for geons also relies on the U-tube
\cite{amalfi}. In the present context, then, it seems important to test the
causal continuity of the U-tube.

In order to use our Morse technology we compactify the cobordism by 
adding a point at spatial infinity at every spatial hypersurface. This creates a 
cobordism between $S^3$ and $P\# P^*$. Then we close off the initial boundary by 
capping it with a $B^4$. This produces a cobordism $\M$ between $\emptyset$ and 
$P\# P^*$ which is $B^4$ with a U-tube of prime $P$.

The question we ask is whether the U-tube cobordism $\M$, admits Morse
functions with $m_1=m_3=0$. In order to do this we first calculate the
Euler characteristic $\chi( \M)$ and then employ equation (\ref{eqn:morse})
which relates it to the $m_{\lambda}$'s.  

\begin{figure}[t]
\centerline{\epsfig{file=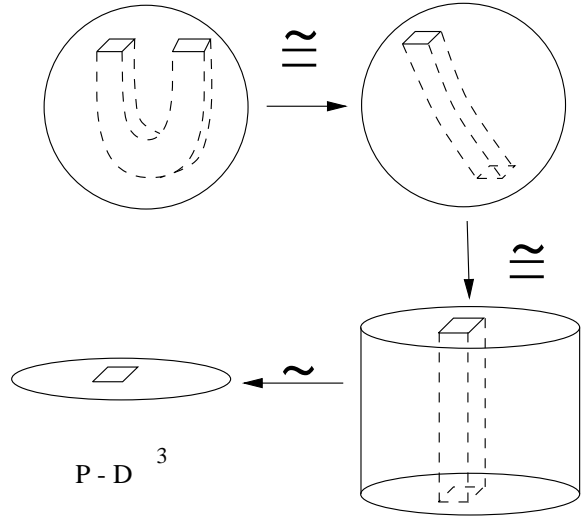,clip=,width=3in}}
\caption{The U-tube pair creation of the prime $P$ is homotopic to $ P -
\D^3$.}
\label{fig:deformed}
\end{figure}

Now, we can unbend the U-tube till it is straight (figure (\ref{fig:deformed}))
to see that $\M\cong I \times (P\# \B^3) \sim P \#
\B^3 \cong P - \D^3$ ( where $\cong$ implies diffeomorphic and $\sim$
homotopy equivalence) and so $\chi(\M) = \chi(P-\D^3)$.

We now use the Mayer-Vietoris sequence for homology groups \cite{rotman}, 
\begin{equation} 
        \cdots \rightarrow H_k(X_1 \cap X_2)\rightarrow H_k(X_1) \oplus
        H_k(X_2) \rightarrow H_k(X) \rightarrow H_{k-1}(X_1 \cap X_2)
        \rightarrow \cdots,
\end{equation}
where $X_1$ and $X_2$ are subspaces of $X$ with $X= int(X_1) \bigcup
int(X_2)$.  Choose $X_1\cong P-\D^3$ and $X_2\cong \B^3$ such that $X = int (X_1)
\bigcup int(X_2) = P$ and $X_1\cap X_2\cong S^2 \times I\sim S^2$.  On
substitution, the above sequence breaks up into the two long exact
sequences,
\begin{equation} 
\label{eqn:sequence}
        0 \rightarrow H_3(P-\D^3){\buildrel {\alpha} \over \rightarrow}
        H_3(P) {\buildrel {\rm D} \over \rightarrow} H_2(S^2){\buildrel
        {\beta} \over \rightarrow} H_2(P-\D^3){\buildrel {\delta} \over
        \rightarrow} H_2(P) \rightarrow 0,
\end{equation} 
and
\begin{equation}
\label{eqn:secseq}
        0 {\buildrel a \over \rightarrow} H_1(P-\D^3) {\buildrel b \over
        \rightarrow} H_1(P) {\buildrel c \over \rightarrow} H_0(S^2)
        {\buildrel d \over \rightarrow} H_0(P-\D^3) \oplus H_0(\D^3)
        {\buildrel e \over \rightarrow} H_0(P) {\buildrel f \over
        \rightarrow}0.
\end{equation}

Let us first examine the map ${\rm D}:H_3(P) \rightarrow H_2(S^2)$ in
(\ref{eqn:sequence}). For an $n$-dimensional space $X=X_1 \bigcup X_2$, each
$n$-cycle $z$ in $X$ is homologous to a cycle of the form ${\gamma}_1 +
{\gamma}_2$ where ${\gamma}_i$ is an $n$-cycle in $X_i$. Moreover, if $D:
H_\lambda(X_1 \bigcup X_2) \rightarrow H_{\lambda-1}(X_1 \cap X_2)$ is the
connecting homomorphism in the Mayer-Vietoris sequence, then $D(cls z)=
D(cls({\gamma}_1 + {\gamma}_2))=cls(\pd {\gamma}_1)$ (Lemma 6.19 in
\cite{rotman}).

Now, $H_3(P)=\Z$ and $H_2(S^2)=\Z$. Let $cls z$ be the generator of
$H_3(P)$. From the above, $D(cls z)=D(cls({\gamma}_1 +
{\gamma}_2))=cls(\pd {\gamma}_1)$, where ${\gamma}_1$ is a 3-cycle in
$P-\D^3$ and ${\gamma}_2$ one in $\B^3$.  Remembering that $P=(P-\D^3) \bigcup
\B^3$ is a closed manifold, the only non-trivial 3-cycle is one that fully
triangulates $P$. This means that $\pd {\gamma}_1$ is a non-trivial
2-cycle in $\pd(P-\D^3)\sim (P-\D^3) \cap \B^3 \sim S^2$ and hence $cls\pd
{\gamma}_1$ is the generator of $H_2(S^2)$. Thus $\rm D$ maps the generator
of $H_3(P)=\Z$ to the generator of $H_2(S^2)=\Z$ which implies that it is
an isomorphism.

Since $\rm D$ is an isomorphism, $ker (\rm D) = 0 = im
({\alpha})$. ${\alpha}$ being a $1-1$ map, $ H_3(P-\D^3)=0$.  Next, $ker
(\beta)=im ({\rm D})=H_2(S^2)$.  Hence $im (\beta)=0= ker(\delta)$. Thus,
$\delta$ which is an onto map is also $1-1$ $\Rightarrow$ $\delta$ is an
isomorphism, or $H_2(P-\D^3)= H_2(P)$.

From (~\ref{eqn:secseq}), using $H_0(X)=\Z$ for $X$ connected, we see that
$ker(e)=\Z=im(d) \Rightarrow$ $d$ is onto and hence $1-1$. Thus,
$ker(d)=0=im(c) \Rightarrow$ that $ker(c)=H_1(P)=im(b)$. This implies that
$b$ is onto and also being $1-1$ is an isomorphism. Thus $H_1(P-D^3) \cong
H_1(P)$.

Summarising, we have,
\begin{eqnarray} 
H_\lambda(P-\D^3)& = & H_\lambda(P) \; \; {\rm for} \;\lambda =0,1,2 \\
                  &= & 0 \; \; {\rm for} \; \lambda  \geq 3.
\end{eqnarray}

Thus, the first three Betti numbers of $\M$: $\beta_0(\M), \beta_1(\M),
\beta_2(\M)$ are the same as those for $P$. Since $P$ is a closed
3-manifold, $\chi (P) =0$, and $\beta_3(P)=1$ and therefore $\chi(\M) =
\chi(P) + 1 = 1$.

From the Morse inequalities we have $m_0\ge 1$ and $m_4 \ge 0$.  Using this
along with the relation (\ref{eqn:morse}) we see that
\begin{equation}
\label{eqn:bds} 
        m_1 + m_3 - m_2 \ge 0
\end{equation} 
Equation (\ref{eqn:bds}) implies that either (a)  $m_1$ or $m_3$ (or both)
are nonzero, or, (b) $m_1=m_2=m_3=0$.

From our earlier comments on the special role played by the big-bang and
big-crunch topologies it seems that (b) must be ruled out since otherwise
there would be no topology change apart from the big-bang creation of an
$S^3$ from nothing. A systematic argument leading to this conclusion
employs the following theorem due to Reeb \cite{hirsch}:

\begin{thm}
\label{thm:reeb}
If $\M$ is a compact $n$-dimensional manifold without boundary, admitting a
Morse function $f:\M \rightarrow [0,1]$ with only two critical points, then
$\M$ is homeomorphic to $S^n$.
\end{thm}

Using this, we now show that (b) leads to a contradiction. First, this
implies that $m_0 = 1$ and $m_4 = 0$. Then, consider the double of $\M$,
the manifold ${\N} = {\M} \bigcup {\bM}$ where $\bM$ is a time-reversed
copy of $\M$ and the union is taken by identifying the boundaries
in the obvious way. 
If $\bar f$ is the time-reversed Morse function on $\bM$ then the
number of index $\lambda$ critical points, ${\bm}_\lambda$ of $\bar f$ are
related to the $m_\lambda$ by $m_\lambda= {\bm}_{n-\lambda}$. We can extend the
Morse function $f$ on $\M$ to some $\F$ on $\N$ as follows,
\begin{eqnarray} 
\F|_{\M} & = & f \\ 
\F|_{\bM}&= & {\bar f}. \\ 
\end{eqnarray} 
$\F$ will therefore have exactly twice the total number of critical points
that $f$ has, and the number of index $\lambda$ points of $\F$ are given by
${\mu}_\lambda= m_\lambda + {\bm}_\lambda= m_\lambda + m_{n-\lambda}$ so
that $\mu_\lambda = \mu_{n-\lambda}$.  
Then $\mu_0 =\mu_4= 1$, $\mu_1=\mu_2=\mu_3=0$ and so $\F$ possesses only two critical
points, one of index $0$ and the other of index $4$. Since $\N$ is a closed
manifold, Theorem \ref{thm:reeb} implies that $\N$ is homeomorphic to
$S^4$, which is clearly false, i.e., (b) is incorrect.

Thus, from (a) we see that {\it any} Morse function $f$ on $\M$ must
possess critical points of index 1 or 3. This means therefore that any
spacetime ($\M, \g$) where $\M$ is a {\it generic} U-tube cobordism in
which an arbitrary prime $P$ is pair-produced will have causal
discontinuities.
Notice that we can choose the prime manifold $P$ to be such that the Betti
numbers of the cobordism are zero, except for $\beta_0$ and $\beta_4$.  For
example, $P= RP^3$. This, then, is an example where the bounds of the Morse
inequalities cannot be realised.

The implications of this result are not very favourable to the particle
picture of primes. It seems that either the picture we have been building
here in which causally discontinuous histories are suppressed in the SOH
fails in some way or the restoration of the spin-statistics correlation for
geons is an illusion (the kinematical calculations of \cite{fay} would
remain true but the more dynamical considerations of causal continuity
would reveal the amplitudes considered to be negligible.) We discuss some
possible ways out in the final section.

\section{Conclusions}
\label{sec:concl}

We have described a rather natural framework for considering topology
change within the SOH for quantum gravity based on Morse theory.  Two
key conjectures lead to the proposal that only causally continuous
cobordisms be included in the Sum and that these are identified with
Morse metrics with no index 1 or $n-1$ points. The Lickorish-Wallace
theorem on surgery on 3-manifolds together with the Borde-Sorkin
conjecture means that any topology changing transition in
3+1-dimensions is achievable by a causally continuous cobordism. The
higher dimensional statement is not known.

        We have shown that the black hole pair production instanton
$S^2\times S^2$ admits causally continuous Morse metrics whereas the
``U-tube'' cobordism for pair production of topological geons of any sort
is necessarily causally discontinuous.

The result on the black hole pair production instanton cobordism fits
in well with the conjectures. However, the topological geon U-tube
pair production cobordism calculation is a serious setback. It is hard
to see how to rescue the spin-statistics theorem for lens spaces if
the U-tube cobordism is indeed suppressed because it cannot support
causally continuous histories. It seems to be {\it the} canonical
pair-creation cobordism and the proof of the theorem rests heavily on
its properties.  Moreover the more general rules proposed by Sorkin
\cite{amalfi} that would lead to a spin-statistics correlation for all
geons also rely on cobordisms that contain U-tubes and these would
also be in jeopardy.

	This might mean that the notion of primes as particles does not
survive with topology change. The causal continuity of the single prime
creation and the causal discontinuity of the U-tube cobordism can then be
regarded as a manifestation of this problem. However, since an important and
physically appealing motivation for topology change comes from the study of
primes as particles \cite{geons} \cite{amalfi}, we suggest here that this
is not the case.

A possible resolution that might save the geon spin-statistics result, is
that there must be a weakness in the sequence of conjectures to which we
have drawn attention and which form the framework in which causal
continuity becomes so central.  The Borde-Sorkin conjecture --- that a
Morse metric is causally continuous iff it contains no index 1 or $(n-1)$
points --- seems to be the most solid. Work on a proof is currently
underway \cite{inprep1}. The Sorkin conjecture that infinite
energy/particle production would occur in a Morse spacetime iff it
contained a causal discontinuity seems plausible but would need to be
verified by more examples than the 1+1 dimensional trousers and yarmulke
studied so far.  In particular, the first example of a causally continuous
spacetime that is not the yarmulke occurs in 3+1 dimensions. Work on this
second conjecture will be easier once the first is proved since then simple
examples of causally continuous metrics can be written down using the Morse
construction.  Then finally, there is the idea that the singular behaviour
of quantum fields on a causally discontinuous background is a signal that
it is infinitely suppressed in the SOH. What one means by this is the
following. Consider a scalar field minimally coupled to gravity. The path
integral is
\begin{equation} 
  \sum_{{\rm all \; topologies}} {\int} [ \dd g] [\dd \phi] {\exp}^{i \int
  \sqrt{-g}R + i \int \sqrt{-g} {(\pd \phi)}^2},         
\end{equation} 
(where we have omitted the explicit and important statement about
boundary
conditions). We may integrate out the scalar field degrees of freedom,
i.e., 
\begin{equation} 
{\int} [\dd \phi] {\exp}^{i \int \sqrt{-g} {(\pd \phi)}^2}= F[g]. 
\end{equation} 
The functional $F[g]$ which is the path integral for a scalar field in a
fixed background can now be regarded as an overall weight in the path
integral over metrics,
\begin{equation} 
  \sum_{{\rm all \; topologies}} {\int} [\dd g] F[g] {\exp}^{i \int
  \sqrt{-g}R}. 
\end{equation}
The idea is that $F[g]$ is zero if $g$ is causally discontinuous. 

Perhaps, however, all the conjectures do hold at the continuum level and
the simplest loophole of all is that the SOH should be defined
fundamentally as a sum over whatever discrete structure will prove to
underly the differentiable manifold of general relativity. If it is a
causal set then all quantities calculated will be regulated.  The
elimination altogether of the causally discontinuous cobordisms would then
be too severe a truncation, and even if they are still suppressed, they 
might give a non-trivial contribution.
\bigskip

\noindent{\bf{Acknowledgements}} We would like to thank Rafael Sorkin for
motivating this work and for several clarifying discussions. We also thank
A. Chamblin, G. Gibbons, R. Penrose, Siddhartha Sen, P. Tod and
N. Woodhouse for helpful discussions. S.Surya would also like to thank
S.M. Vaidya for help with procuring references. F. Dowker is supported in
part by an EPSRC Advanced Fellowship and thanks the Department of Applied
Maths and Theoretical Physics at University of Cambridge for hospitality
during part of this work. Sumati Surya was supported in part by a Syracuse
University Fellowship and by Inter University Centre for Astronomy and
Astrophysics, Pune.

\bibliographystyle{unsrt}

\bibliography{refs}
\end{document}